\documentclass{article}
\pdfoutput=1

\usepackage{amsmath,amssymb,bm,cite}
\def\rd{\mathrm{d}}
\def\ri{\mathrm{i}}
\def\id{\mathbf{1}}

\def\fl{\mathrm{fl}}
\begin{document}
\begin{titlepage}

\noindent {\large \textbf{Closedness of orbits in a space with SU(2) Poisson structure}}

\vspace{2\baselineskip}
\begin{center}

Amir~H.~Fatollahi~{\footnote{fath@alzahra.ac.ir}}
\vspace{0.5\baselineskip}

Ahmad~Shariati~{\footnote{shariati@mailaps.org}}
\vspace{0.5\baselineskip}

Mohammad~Khorrami~{\footnote{mamwad@mailaps.org}}

\vspace{2\baselineskip}
{\it Department of Physics, Alzahra University, Tehran 19938-93973, Iran}

\end{center}

\vspace{1\baselineskip}

\begin{abstract}
\noindent The closedness of orbits of central forces is addressed
in a three dimensional space in which the Poisson bracket among the
coordinates is that of the SU(2) Lie algebra.
In particular it is shown that among problems with spherically symmetric
potential energies, it is only the Kepler problem
for which all of the bounded orbits are closed. In analogy
with the case of the ordinary space, a conserved vector (apart from
the angular momentum) is explicitly constructed, which is responsible
for the orbits being closed. This is the analog of
the Laplace-Runge-Lenz vector. The algebra of the constants of the motion
is also worked out.
\end{abstract}

\vspace{2\baselineskip}


\end{titlepage}
\section{Introduction}
There has been great interest in the study
of physical theories on noncommutative spaces.
The motivation is partly the natural appearance of noncommutative spaces
in some areas of physics, for example in the string theory 
(\cite{doplicher1,doplicher2,madore1,madore2,madore3}, for example). In particular, it has been 
understood that the canonical relation
\begin{equation}\label{10.1}
[\hat x_a,\hat x_b]=\ri \,\theta_{a\, b}\,\id,
\end{equation}
with $\theta_{a\, b}$ being an antisymmetric constant tensor,
describes the commutators of the longitudinal directions of $D$-branes 
in the presence of a constant $B$-field background 
\cite{9908142,99-2,99-3,99-4}. The implications of this kind of 
coordinates have been extensively studied \cite{reviewnc1,reviewnc2}.

One natural direction to extend studies on noncommutative spaces is to
consider cases when the commutators of the coordinates are not
constants. Examples of this kind are the $\kappa$-Poincar\'{e} algebra 
\cite{majid,ruegg1,ruegg2,ruegg3,amelino1,amelino2,kappa1,kappa2,kappa3}, the noncommutative cylinder
and the $q$-deformed plane \cite{chai1,chai2}, 
and linear noncommutativity of the Lie algebra type
\cite{wess,sasak1,sasak2,sasak3}. In the latter the dimensionless spatial
position operators satisfy:
\begin{equation}\label{10.2}
[\hat{x}_a,\hat{x}_b]= f^c{}_{a\, b}\,\hat{x}_c,
\end{equation}
where $f^c{}_{a\,b}$'s are structure constants of a Lie algebra.
One example of such algebras is the algebra SO(3), or SU(2). A special
case of this, is the so-called fuzzy sphere \cite{madore,presnaj1,presnaj2,presnaj3},
where an irreducible representation of the position operators is
used which makes the Casimir of the algebra a multiple of the
identity operator (a constant, hence the name sphere).

In \cite{0612013,fakE1,fakE2,spga1,spga2} a model was introduced in which the
representation was not restricted to an irreducible one, but all of
the irreducible representations were used; see also \cite{jablam}. 
More specifically, the regular representation of the group was considered, 
which contains all irreducible representations. As a consequence in such models 
one is dealing with the whole space, rather than a sub-space like the case of the
fuzzy sphere. In \cite{0612013,fakE1,fakE2,spga1,spga2} basic
ingredients for calculus on a linear fuzzy space, as well as the field theoretic
aspects on such a space, were studied in details.
It is observed that models based on
Lie algebra type noncommutativity enjoy three features:
\begin{itemize}
\item They are free from any ultraviolet divergences if the group
is compact. (In fact this has nothing to do with the noncommutativity,
it is a result of the momentum space being compact.)
\item The momentum conservation is modified, in the
sense that the vector addition is replaced by some  non-Abelian
operation \cite{pal,0612013}.
\item In the transition amplitudes only the so-called planar
graphs contribute.
\end{itemize}
In \cite{fsk} the quantum mechanics
on a space with SU(2) fuzziness was examined. In particular,
the commutation relations of the position and momentum
operators corresponding to spaces with Lie-algebra
noncommutativity in the configuration space were studied.
The consequences of the Lie type noncommutativity of space on
thermodynamical properties have been explored in \cite{shin1,shin2,fsmjmp}.

The classical motion on noncommutative space
has attracted interests as well \cite{miao,silva}.
In particular, the central force problems on space-times
with canonical and linear noncommutativity and their
observational consequences have been the subject of different
research works \cite{levia,zhang,kapoor,dennis,romero,mirza}.
In \cite{kfs} the classical mechanics defined on a space with
SU(2) fuzziness was studied. In particular, the Poisson structure
induced by noncommutativity of SU(2) type was investigated, for
either the Cartesian or Euler parameterization of SU(2) group.

In \cite{andalib} the central force problem on a space with SU(2)
Poisson structure was revisited. In particular, it was shown that on such a space
it is only the Kepler potential energy, as a single-term power-law one,
for which all of nearly circular orbits are closed.
The others, if any, should satisfy a differential equation, and certainly
are not single-term power law. For the case of the Kepler potential energy, 
it was shown by explicitly solving the differential equation for the orbit,
that all bounded orbits are closed.

The purpose of the present work is to continue the study of central force
problem on space with SU(2) Poisson structure.
Here it is shown that among all spherically symmetric
potential energies, power-law or not, it is only the
Kepler problem for which all of the bounded orbits are closed.
This result is in contrast with the case of ordinary space, in which
the Bertrand's theorem \cite{goldstein} states that there are
two potential energies for which all bounded orbits are closed,
which correspond to the harmonic oscillator and the Kepler problem. 
It is seen that while the bounded orbits of the Kepler problem remain 
closed even in the case of SU(2) noncommutativity, that is not the case 
for the harmonic oscillator problem. In analogy with the case of 
the ordinary space, there exists a conserved vector (apart from 
the angular momentum), which is responsible for the bounded orbits 
of the Kepler problem be closed. This vector, the noncommutative 
analog of the Laplace-Runge-Lenz vector, is explicitly constructed. 
The algebra of the constants of the motion is also worked out.

The scheme of the rest of this paper is the following. In section 2,
a short review of the construction in \cite{kfs} is presented.
In section 3, an analysis of the nearly circular orbits is performed,
which results in a differential equation for potential energies
for which the nearly circular orbits are closed. Comparing
the result with the case of commutative space, it is shown that
necessary conditions for all bounded orbits being closed, is
fulfilled by only two potential energies, those of the harmonic oscillator
and the Kepler problem. The harmonic oscillator is ruled out,
by an explicit counter example, showing that some bounded orbit
is not closed, and it is explicitly shown that the all of the bounded orbits 
of the Kepler problem are closed. In section 4, the analog of 
the Laplace-Runge-Lenz vector is explicitly constructed
and the algebra of the constants of motion is worked out.
Section 5 is devoted to the concluding remarks.
\section{Basic notions}
The classical dynamics on a space whose Poisson structure is
originated from a Lie algebra is given in \cite{kfs}. To make this paper
self contained, a short review of the construction is presented below.

Denote the members of a basis for the left-invariant vector fields
corresponding to the group $\mathbb{G}$ by $\hat{x}_a$'s. These fields satisfy
the commutation relation (\ref{10.2}), with the structure constants
of the Lie algebra corresponding to $\mathbb{G}$. The members of this basis
would correspond to the quantum mechanical (operator form) counterparts
of the classical spatial coordinates of the system.
The group elements of $\mathbb{G}$ are parametrized by the coordinates $k^a$ as:
\begin{equation}\label{10.3}
U(\bm{k}):=[\fl({k}^a\,\hat{x}_a)][U(\bm{0})],
\end{equation}
where $\fl(v)$ is the flow of the vector field $v$.
These coordinates would play the role of the conjugate momenta of
$\hat{x}_a$'s. The coordinates and the momenta in their operator forms
(denoted by hat) would satisfy the following relations
\begin{align}\label{10.4}
[\hat{k}^a,\hat{k}^b]&=0,\\
\label{10.5} [\hat{x}_a,\hat{k}^b]&=\hat{x}_a{}^b,
\end{align}
with ${x}_a{}^b$'s being scalar functions of $\bm{k}$,
having the property
\begin{equation}\label{10.6}
{x}_a{}^b(\bm{k}=\bm{0})=\delta_a^b.
\end{equation}
Accordingly, the operator forms of $\hat{x}_a$'s in the
$k$-basis are
\begin{equation}\label{10.7}
\hat{x}_a\to {x}_a{}^b\,\frac{\partial~}{\partial k^b}.
\end{equation}
There are also the right-invariant vector fields
whose basis ($\hat{x}_a^{\mathrm{R}}$'s) satisfies
\begin{align}\label{10.8}
[\hat{x}_a^{\mathrm{R}},\hat{x}_b^{\mathrm{R}}]&=
-f^c{}_{a\,b}\,\hat{x}_c^{\mathrm{R}},\\
\label{10.9} [\hat{x}_a^{\mathrm{R}},\hat{x}_b]&=0.
\end{align}
Using these, one defines the vector field $\hat{J}_a$ through
\begin{equation}\label{10.10}
\hat{J}_a:=\hat{x}_a-\hat{x}_a^{\mathrm{R}}.
\end{equation}
Using (\ref{10.9}) and the definitions of the left- and right-actions,
it is found that $J_a$'s are the generators of similarity transformation
(adjoint action):
\begin{equation}\label{10.11}
[\fl(\alpha^a\,\hat{J}_a)][U(\bm{k})]=[U(-\bm{\alpha})]\,[U(\bm{k})]\,[U(\bm{\alpha})].
\end{equation}
One also finds the following commutation relations.
\begin{align}\label{10.12}
[\hat{J}_a,\hat{J}_b]&=f^c{}_{a\,b}\,\hat{J}_c,\\
\label{10.13} [\hat{J}_a,\hat{x}_b]&=f^c{}_{a\,b}\,\hat{x}_c,\\
\label{10.14} [\hat{k}^c,\hat{J}_a]&=f^c{}_{a\,b}\,\hat{k}^b.
\end{align}
In the case of the group SU(2), the structure constants $f^c{}_{a\,b}$ 
are the same as $\varepsilon^c{}_{a\,b}$ (the components of 
the Levi-Civita tensor), and the above commutations show that 
$\hat J_a$'s are the generators of rotation (in the group manifold), 
and thus represent the components of the angular momentum.

To construct the phase space, all that is needed is to change
the commutation relations into Poisson brackets:
\begin{align}\label{10.15}
\{k^a,k^b\}&=0,\\
\label{10.16} \{x_a,k^b\}&=x_a{}^b,\\
\label{10.17} \{x_a,x_b\}&=f^c{}_{a\,b}\,x_c,\\
\label{10.18} \{J_a,x_b\}&=f^c{}_{a\,b}\,x_c,\\
\label{10.19} \{k^c,J_a\}&=f^c{}_{a\,b}\,k^b,\\
\label{10.20} \{J_a,J_b\}&=f^c{}_{a\,b}\,J_c.
\end{align}
Here $J_a$'s and $x_a$'s have the dimension of action, while
$k^a$'s are dimensionless. One can, of course, use a parameter
$\lambda$ of the dimension of the linear momentum inverse, to
construct $(\lambda\,x_a)$ and $(\lambda^{-1}\,k^a)$
with the dimension of length and linear momentum, respectively.
The commutative limit is achieved when one sends $\lambda$ to zero
while keeping $(\lambda\,x_a)$ and $(\lambda^{-1}\,k^a)$ constant.
One notes that $x_a$'s and $k^a$'s are independent variables, and
other variables can be expressed in terms of these. So that among
the Poisson brackets (\ref{10.15}) to (\ref{10.20}), only
(\ref{10.15}), (\ref{10.16}), and (\ref{10.17}) are
independent. All others can be derived from these.
\subsection{The SU(2) setup}
For the special case of the group SU(2), where the structure constants
$f^c{}_{a\,b}$ are $\varepsilon^c{}_{a\,b}$, the independent Poisson brackets
(\ref{10.15}), (\ref{10.16}), and (\ref{10.17}) are in fact the
Poisson structure of a rigid rotator, in which the angular
momentum and the rotation vector have been replaced by
$\bm{x}$ and $\bm{k}$, respectively, that is, the roles of
position and momenta have been interchanged.
One can use the Euler parameters (instead of the rotation
vector $\bm{k}$) as the coordinates of the group.
The Euler parameters $(\phi,\theta,\psi)$ are defined
through
\begin{equation}\label{10.21}
[\exp(\phi\,T_3)]\,[\exp(\theta\,T_2)]\,[\exp(\psi\,T_3)]:=[\exp(k^a\,T_a)],
\end{equation}
where $T_a$'s are the generators of SU(2) satisfying the
commutation relation
\begin{equation}\label{10.22}
[T_a,T_b]=\varepsilon^c{}_{a\,b}\,T_c.
\end{equation}
It is to be noted that here, the Euler parameters are coordinates of
the momentum space, unlike the case of the rigid rotator where they are
the coordinates of the configuration space. Here the coordinates
of the configuration space, conjugate to the Euler parameters, are
denoted by $X_\phi$, $X_\theta$, and $X_\psi$.
So these, added to the Euler parameters, are six canonical coordinates
of the phase space, meaning that the only non-vanishing Poisson brackets
between them are
\begin{align}\label{10.23}
\{X_\phi,\phi\}&=1,\\
\label{10.24} \{X_\theta,\theta\}&=1,\\
\label{10.25} \{X_\psi,\psi\}&=1.
\end{align}
One can then express the coordinates of the position
(the left-invariant vector fields) as \cite{kfs}
\begin{align}\label{10.26}
x_1&=\left[-\frac{\cos\psi}{\sin\theta}\,X_\phi+\sin\psi\,X_\theta+
\frac{\cos\psi\,\cos\theta}{\sin\theta}\,X_\psi\right],\\
\label{10.27}
x_2&=\left[\frac{\sin\psi}{\sin\theta}\,X_\phi+\cos\psi\,X_\theta-
\frac{\sin\psi\,\cos\theta}{\sin\theta}\,X_\psi\right],\\
\label{10.28} x_3&=X_\psi.
\end{align}
Similarly the angular momentum components are found as follows \cite{kfs}.
\begin{align}
\label{10.29}
J_1&=\frac{\cos\phi\,\cos\theta-\cos\psi}{\sin\theta}\,X_\phi+
(\sin\phi+\sin\psi)\,X_\theta \cr &+
\frac{-\cos\phi+\cos\psi\,\cos\theta}{\sin\theta}\,X_\psi,\\
\label{10.30}
J_2&=\frac{\sin\phi\,\cos\theta+\sin\psi}{\sin\theta}\,X_\phi
+(-\cos\phi+\cos\psi)\,X_\theta\cr &+
\frac{-\sin\phi-\sin\psi\,\cos\theta}{\sin\theta}\,X_\psi,\\
\label{10.31} J_3&=-X_\phi+X_\psi.
\end{align}
One also has \cite{kfs}
\begin{equation}\label{10.32}
\cos\frac{ k}{2}=\cos\frac{\theta}{2}\,\cos\frac{\phi+\psi}{2},
\end{equation}
where
\begin{equation}\label{10.33}
k:=\sqrt{\bm{k}\cdot\bm{k}},
\end{equation}
and
\begin{equation}\label{10.34}
\bm{y}\cdot\bm{z}:=\delta_{a\,b}\,y^a\,z^b.
\end{equation}

In the case of motion under a central force, the Poisson
brackets of Hamiltonian $H$ with $J_a$'s vanish. A Hamiltonian
which is a function of only $(\bm{k}\cdot\bm{k})$ and
$(\bm{x}\cdot\bm{x})$ is clearly so. For such a system, $H$,
$\bm{J}\cdot\bm{J}$, and one of the components of $\bm{J}$
(say $J_3$) are involutive constants of motion, hence
any SU(2)-invariant classical system is integrable.
As $\bm{J}$ is a constant vector, one can choose the axes
so that the third axis is parallel to this vector:
\begin{align}\label{10.35}
J_1&=0,\\
\label{10.36} J_2&=0.
\end{align}
Using these, and assuming that $J_3$ is non-vanishing, one arrives at
\begin{align}\label{10.37}
\phi+\psi&=0,\\
\label{10.38} X_\phi+X_\psi&=0.
\end{align}
Defining the new variables
\begin{align}\label{10.39}
\chi&:=\frac{\psi-\phi}{2},\\
\label{10.40} J&:=-X_\phi+X_\psi,
\end{align}
it is seen that $J$ is the conjugate to the momentum coordinate $\chi$, so
that among the four coordinates $(J,X_\theta,\chi,\theta)$, the only
non-vanishing Poisson brackets are
\begin{align}
\{X_\theta,\theta\}&=1,\nonumber\tag{\ref{10.24}}\\
\label{10.41} \{J,\chi\}&=1.
\end{align}
In terms of these,
\begin{align}\label{10.42}
x_1&=\left(\frac{J}{2}\,\frac{1+\cos\theta}{\sin\theta}\,\cos\chi+
X_\theta\,\sin\chi\right),\\
\label{10.43}
x_2&=\left(-\frac{J}{2}\,\frac{1+\cos\theta}{\sin\theta}\,\sin\chi+
X_\theta\,\cos\chi\right),\\
\label{10.44}
x_3&=\frac{J}{2},\\
\label{10.45} \bm{x}\cdot\bm{x}&=\left[X_\theta^2+
\frac{J^2}{4}\,\left(1+\cot^2\frac{\theta}{2}\right)\right],\\
\label{10.46} \cos\frac{{k}}{2}&=\cos\frac{\theta}{2}.
\end{align}
Defining the polar coordinates ($\rho, \alpha$) in the $(x_1, x_2)$ plane, and also $u$:
\begin{align}\label{10.47}
x_1&=:\rho\,\cos\alpha,\\
\label{10.48} x_2&=:\rho\,\sin\alpha,\\
\label{10.49} \frac{1}{2}\,\cot\frac{\theta}{2}&=:u,
\end{align}
one arrives at
\begin{align}\label{10.50}
\bm{x}\cdot\bm{x}&=\rho^2+\frac{J^2}{4},\\
\label{10.51} \rho^2&=X_\theta^2+J^2\,u^2,\\
\label{10.52} \alpha&=-\chi+\tan^{-1}\frac{X_\theta}{J\,u}.
\end{align}
One then has
\begin{equation}\label{10.53}
\bm{x}\cdot\bm{x}=X_\theta^2+
J^2\,\left(\frac{1}{4}+u^2\right).
\end{equation}
In analogy with the case of commutative space,
one can proceed to derive the equation for the path, in terms
of $\rho$ and $\alpha$. The first-order path equation
is found to be \cite{kfs}:
\begin{equation}\label{10.54}
\frac{1}{u^2}=J^2\,\left[\frac{1}{\rho^2}+\frac{1}{\rho^4}
\left(\frac{\rd\rho}{\rd\alpha}\right)^2\,\right].
\end{equation}
Using (\ref{10.49}) and (\ref{10.46}), $u$ can be expressed
in terms of the kinetic energy.
Following \cite{0612013,fakE1,fakE2,kfs}, the kinetic energy for
a particle of mass $m$ is taken to be
\begin{equation}\label{10.55}
K=\frac{4}{m}\,\left(1-\cos\frac{k}{2}\right).
\end{equation}
It is SU(2)-invariant, monotonically increasing for $0\leq k\leq 2\,\pi$,
and tends to $(2\,m)^{-1}\,k^2$ for small values of $k$, which is
the commutative limit. Expressing $u$ in terms of $K$, one has
\begin{equation}\label{10.56}
\frac{1}{u^2}=4\,\left[\left(1-\frac{m}{4}\,K\right)^{-2}-1\right].
\end{equation}
Assuming the usual form for the Hamiltonian as
\begin{align}\label{10.57}
H&=K+V(r),\\
\label{10.58}
\sqrt{\bm{x}\cdot\bm{x}}&=:r,
\end{align}
the path equation (\ref{10.54}) is rewritten as
\begin{equation}\label{10.59}
\frac{J^2}{4}\,
\left[\frac{1}{\rho^2}+\frac{1}{\rho^4}\left(\frac{\rd\rho}{\rd\alpha}\right)^2\right]=
\left\{1-\frac{m}{4}\,[E-V(r)]\right\}^{-2}-1,
\end{equation}
in which
\begin{align}\label{10.60}
r&=\sqrt{\rho^2+x_3^2},\nonumber\\
&=\sqrt{\rho^2+\frac{J^2}{4}}.
\end{align}

\section{Closed or not}
As mentioned earlier, it has been shown that on a space with SU(2)
Poisson structure there is only one power-law potential energy
for which all of nearly circular orbits are closed \cite{andalib}.
Others, if any, should satisfy a differential equation, and are
not power-laws. Here it is shown that among spherically symmetric
potential energies, there is only one potential energy for which
all bounded orbits are closed, and that potential energy
is the potential energy corresponding to the Kepler problem.
First, a criterion is obtained for spherically symmetric
potential energies with the property that the nearly circular orbits
corresponding to them be closed. It is then shown that among these,
only two potential energies satisfy necessary conditions for all bounded
orbits being closed. One of these is a noncommutative analog of
the harmonic oscillator potential energy (though it is not
a power law in the noncommutative case), the other that of the
Kepler problem, where the potential energy is proportional to
the inverse of the distance from the origin. A counter example is
presented to show that for the analog of the harmonic oscillator,
not all bounded orbits are closed. For the Kepler problem, it is
explicitly shown that all bounded orbits are closed.

\subsection{Closedness criteria}
Denoting the angular period of $r$ by $\mathcal{T}$, one has
\begin{equation}\label{10.61}
\mathcal{T}=2\,\int_{r_\mathrm{min}}^{r_\mathrm{max}}\rd r\,\alpha(r).
\end{equation}
The equations used to calculate $\mathcal{T}$, are (\ref{10.59}) and (\ref{10.60}).
Defining
$\mathcal{E}$, $\mathcal{V}$, $w$, and $s$ through
\begin{align}\label{10.62}
\mathcal{E}&:=\frac{m\,E}{4},\\
\label{10.63} \mathcal{V}&:=\frac{m\,V}{4},\\
\label{10.64} \frac{J}{2\,\sinh w}&:=\rho,\\
\label{10.65} s&:=\ln\frac{2\,r}{J},
\end{align}
One arrives at
\begin{align}\label{10.66}
\coth w&=\exp(s),\\ \label{10.67}
r\,\frac{\partial}{\partial r}&=\left(\frac{\partial}{\partial s}\right)_{J, E},\nonumber\\
&=-\sinh w\,\cosh w\,\left(\frac{\partial}{\partial w}\right)_{J, E}.
\end{align}
So,
\begin{equation}\label{10.68}
\cosh^2 w\,\left[1+\left(\frac{\rd w}{\rd\alpha}\right)^2\right]=
[1-\mathcal{E}+\mathcal{V}(r)]^{-2}.
\end{equation}
Defining the function $B$ through
\begin{equation}\label{10.69}
B(w)=[1-\mathcal{E}+\mathcal{V}(r)]\,\cosh w,
\end{equation}
it is seen that to have a nearly circular orbit around $w=w_0$, one should have
\begin{align}\label{10.70}
B(w_0)&=1.\\ \label{10.71}
\left(\frac{\partial B}{\partial w}\right)_{J, E}(w_0)&=0.\\ \label{10.72}
\left(\frac{\partial^2 B}{\partial w^2}\right)_{J, E}(w_0)&=q^2.
\end{align}
It is seen then that for nearly circular orbits,
\begin{equation}\label{10.73}
\mathcal{T}=\frac{2\,\pi}{q}.
\end{equation}
Equations (\ref{10.70}) and (\ref{10.71}) then become
\begin{align}\label{10.74}
1&=(\cosh w_0)\,[1-\mathcal{E}+\mathcal{V}(r_0)].\\
\label{10.75} 0&=(\sinh w_0)\,[1-\mathcal{E}+\mathcal{V}(r_0)]
+(\cosh w_0)\,\left(\frac{\partial\mathcal{V}}{\partial w}\right)_{J, E}(w_0),\nonumber\\
&=\tanh w_0-\frac{1}{\sinh w_0}\,\left(\frac{\rd\mathcal{V}}{\rd s}\right)_0,
\end{align}
where the subscript $0$ means that the corresponding quantity has been calculated
at $w_0$. So,
\begin{equation}\label{10.76}
\frac{\sinh^2 w_0}{\cosh w_0}=\left(\frac{\rd\mathcal{V}}{\rd s}\right)_0.
\end{equation}
Finally, (\ref{10.72})  results in
\begin{align}\label{10.77}
q^2&=(\cosh w_0)\,[1-\mathcal{E}+\mathcal{V}(r_0)]\nonumber\\
&\quad+2\,(\sinh w_0)\,\left(\frac{\partial\mathcal{V}}{\partial w}\right)_{J, E}(w_0)
+(\cosh w_0)\,\left(\frac{\partial^2\mathcal{V}}{\partial w^2}\right)_{J, E}(w_0),\nonumber\\
&=1-\frac{2}{\cosh w_0}\,\left(\frac{\rd\mathcal{V}}{\rd s}\right)_0\nonumber\\
&\quad+(\cosh w_0)\,
\left[\frac{1}{(\sinh^2 w_0)\,(\cosh^2 w_0)}\,\left(\frac{\rd^2\mathcal{V}}{\rd s^2}\right)_0\right.
\nonumber\\
&\left.\quad+\frac{2\,\sinh^2 w_0+1}{(\sinh^2 w_0)\,(\cosh^2 w_0)}\,
\left(\frac{\rd\mathcal{V}}{\rd s}\right)_0\right],\nonumber\\
&=1+\frac{1}{(\sinh^2 w_0)\,(\cosh w_0)}\,\left(\frac{\rd^2\mathcal{V}}{\rd s^2}+
\frac{\rd\mathcal{V}}{\rd s}\right)_0,\nonumber\\
&=1+\frac{1}{\cosh^2 w_0}\,\left(\frac{\rd\mathcal{V}}{\rd s}\right)_0^{-1}\,
\left(\frac{\rd^2\mathcal{V}}{\rd s^2}+\frac{\rd\mathcal{V}}{\rd s}\right)_0.
\end{align}
In order that nearly circular orbits be closed, $\mathcal{T}$
should be a rational multiple of $(2\,\pi)$, and as $\mathcal{T}$ is a continuous
function of $E$ and $J$, it should be independent of $E$ and $J$ for such orbits.
So $q$ should be independent of $E$ and $J$ (and hence independent of $w_0$).
So the above equation is a differential equation for $\mathcal{V}$ with constant
$q$, if nearly circular orbits are to be closed \cite{goldstein}. Defining a variable $\Upsilon$,
and using (\ref{10.76}),
\begin{align}\label{10.78}
\Upsilon&=\frac{1}{2}\,\frac{\rd\mathcal{V}}{\rd s}.\\
\label{10.79} \cosh w_0&=(\Upsilon+\sqrt{\Upsilon^2+1})_0,
\end{align}
one can rewrite the differential equation for $\mathcal{V}$ as
\begin{equation}\label{10.80}
\frac{1}{\Upsilon}\,\frac{\rd\Upsilon}{\rd s}=[(q^2-1)\,(\Upsilon+\sqrt{\Upsilon^2+1})^2-1].
\end{equation}
In terms of the variables $c$ and $\Xi$, with
\begin{align}\label{10.81}
c&=q^2-1,\\
\label{10.82} \Upsilon&=\sinh\left(\frac{1}{2}\,\ln\Xi\right),\nonumber\\
&=\frac{1}{2}\,(\Xi^{1/2}-\Xi^{-1/2}),
\end{align}
the differential equation would be
\begin{equation}\label{10.83}
\frac{\Xi+1}{2\,\Xi\,(\Xi-1)}\,\frac{\rd\Xi}{\rd s}=c\,\Xi-1,
\end{equation}
the solution to which is
\begin{equation}\label{10.84}
\exp[2\,(s-\sigma)]=\Xi\,(\Xi-1)^{2/(c-1)}\,(\Xi-c^{-1})^{-(c+1)/(c-1)},
\end{equation}
where $\sigma$ is a constant. This shows that for fixed $c$ (or $q$),
the function $\mathcal{V}$ contains only two constants. One is an additive
constant, which is not important as it can be absorbed in $\mathcal{E}$.
The other ($\sigma$), can be rewritten in terms $\tilde r$:
\begin{equation}\label{10.85}
\sigma=\ln\frac{2\,\tilde r}{J}.
\end{equation}
So $\mathcal{V}$ has the following form (apart from the additive constant).
\begin{equation}\label{10.86}
\mathcal{V}(r)=U_c\left(\frac{r}{\tilde r}\right).
\end{equation}
It is seen from (\ref{10.59}), that if $\mathcal{V}$ is like the above,
then $\mathcal{T}$ depends on only
$\tilde r$, $J$, $\mathcal{E}$, and of course $c$:
\begin{equation}\label{10.87}
\mathcal{T}(m,c,\tilde r, J, E)=\mathfrak{F}(c,\tilde r,J,\mathcal{E}).
\end{equation}
But again (\ref{10.59}) shows that a solution to that equation is
changed into another solution, if one multiplies $J$
and $\tilde r$ by the same constant $\mathfrak{m}$, while keeping
$\mathcal{E}$ constant:
\begin{equation}\label{10.88}
\mathfrak{F}(c,\mathfrak{m}\,\tilde r,\mathfrak{m}\,J,\mathcal{E})=
\mathfrak{F}(c,\tilde r,J,\mathcal{E}).
\end{equation}
Demanding that $\mathcal{T}$ be unchanged when only $J$ is changed,
one arrives at
\begin{equation}\label{10.89}
\mathfrak{F}(c,\mathfrak{m}\,\tilde r,\mathfrak{m}\,J,\mathcal{E})=
\mathfrak{F}(c,\mathfrak{m}\,\tilde r,J,\mathcal{E}).
\end{equation}
A comparison of the last two equations shows that $\mathfrak{F}$
should be independent of its second variable:
\begin{equation}\label{10.90}
\mathcal{T}(m,c,\tilde r,J,E)=\mathfrak{F}(c,J,\mathcal{E}).
\end{equation}
The dimension of $\rho$ is that of the action. A corresponding quantity
with the dimension of length would be $(\lambda\,\rho)$. Keeping this constant,
while sending $\lambda$ to zero, one arrives at the commutative limit.
So the commutative limit corresponds to large $\rho$ (while $J$ remains finite).
It is seen that in this limit, (\ref{10.59}) and (\ref{10.60}) become
\begin{alignat}{2}\label{10.91}
\frac{J^2}{4}\,
\left[\frac{1}{\rho^2}+\frac{1}{\rho^4}\left(\frac{\rd\rho}{\rd\alpha}\right)^2\right]&=
\frac{m}{2}\,[E-V(r)],&\quad&\mbox{commutative limit}\\
\label{10.92} r&=\rho,&\quad&\mbox{commutative limit}
\end{alignat}
which are the equations of orbits in the commutative case,
as expected. So a necessary condition for all bounded orbits
{\em in the noncommutative case} to be closed, is that the potential energy
 in the commutative case be so that all bounded orbits
{\em in the commutative case} be closed. As the angular period should be
independent of $\tilde r$, the only other parameter in $\mathcal{V}$
(namely $c$) should be so that this condition is satisfied.
It is known (Bertrand's theorem \cite{goldstein}), that in the
commutative case there only two spherically symmetric potential energies
with such a property. It is also seen that the commutative limit corresponds to
$\mathcal{V}$ and $\Upsilon$ tending to zero and $\Xi$ tending to 1, so
that (\ref{10.83}) becomes
\begin{equation}\label{10.93}
\frac{\rd\Xi}{\rd s}=(c-1)\,\Xi,\quad\mbox{commutative limit}
\end{equation}
showing that in the commutative limit $\Xi$ and hence $\Upsilon$ and $\mathcal{V}$
are power-laws, again as expected. In fact, it is seen that
\begin{equation}\label{10.94}
V(r)\propto r^{c-1},\quad\mbox{commutative limit}
\end{equation}
showing that closed orbits in the commutative limit correspond to
$c=3$, which is the harmonic oscillator problem; and $c=0$,
which is the Kepler problem. So in the noncommutative case too,
the necessary condition for all bounded orbits be closed is that
$c$ be equal to either $3$ or $0$.
\subsection{The harmonic oscillator problem}
To obtain the angular period $\mathcal{T}$, one considers
\begin{align}\label{10.95}
\frac{\rd\mathcal{V}}{\rd\Xi}&=\frac{\rd\mathcal{V}}{\rd s}\,\frac{\rd s}{\rd\Xi},\nonumber\\
&=\frac{\Xi^{1/2}-\Xi^{-1/2}}{2}\,\left(\frac{1}{\Xi}+\frac{2}{c-1}\,\frac{1}{\Xi-1}-
\frac{c+1}{c-1}\,\frac{1}{\Xi-c^{-1}}\right).
\end{align}
Defining
\begin{equation}\label{10.96}
\xi:=\Xi^{1/2},
\end{equation}
one arrives at
\begin{equation}\label{10.97}
\frac{\rd\mathcal{V}}{\rd\xi}=(\xi^2-1)\,\left(\frac{1}{\xi^2}+\frac{2}{c-1}\,\frac{1}{\xi^2-1}-
\frac{c+1}{c-1}\,\frac{1}{\xi^2-c^{-1}}\right),
\end{equation}
so that
\begin{equation}\label{10.98}
1-\mathcal{E}+\mathcal{V}=\kappa+\frac{1}{\xi}-\frac{c+1}{\sqrt{c}}\,
\begin{cases}\coth^{-1}(\sqrt{c}\,\xi)\\ \tanh^{-1}(\sqrt{c}\,\xi)\end{cases},
\end{equation}
where $\kappa$ is an integration constant. One has
\begin{equation}\label{10.99}
\mathcal{T}=2\,\int_{\xi=\xi_1}^{\xi=\xi_2}\frac{(1-\mathcal{E}+\mathcal{V})\,\exp(-s)\,\rd s}
{[1-\exp(-2\,s)]\,\sqrt{1-\exp(-2\,s)-(1-\mathcal{E}+\mathcal{V})^2}},
\end{equation}
where $\xi_1$ and $\xi_2$ are the roots of the square-root in the denominator of
the integrand. Defining
\begin{align}\label{10.100}
S&:=\exp(-2\,s),\\
\label{10.101} \nu&:=\exp(-2\,\sigma),
\end{align}
one arrives at
\begin{align}\label{10.102}
S&=\nu\,\Xi^{-1}\,(\Xi-1)^{-2/(c-1)}\,(\Xi-c^{-1})^{(c+1)/(c-1)},\\
\label{10.103}
1-\mathcal{E}+\mathcal{V}&=\kappa+\frac{1}{\sqrt{\Xi}}-\frac{c+1}{\sqrt{c}}\,\coth^{-1}(\sqrt{c\,\Xi}),
\qquad c>1,\\
\label{10.104}
\mathcal{T}&=-2\,\int_{\Xi=\Xi_1}^{\Xi=\Xi_2}\frac{1-\mathcal{E}+\mathcal{V}}
{(1-S)\,\sqrt{1-S-(1-\mathcal{E}+\mathcal{V})^2}}\,\frac{\rd\sqrt{S}}{\rd\Xi}\,\rd\Xi.
\end{align}
Evaluating $\mathcal{T}$ numerically for $c=3$, it is seen that
\begin{equation}\label{10.105}
\frac{\mathcal{T}(c=3,\nu=0.25,\kappa=0.87)}{\pi}=0.985,
\end{equation}
as an example. For the orbit to be closed, the right-hand side
should have been equal to one. So for $c=3$, the analog of
the harmonic oscillator problem, the orbits are not closed.
\subsection{The Kepler problem}
For the case $c=0$, the equation relating $w$ to $\alpha$ can
be explicitly solved, to show that all bounded orbits are
in fact closed \cite{andalib}. One has from
(\ref{10.80}), with $c=0$ which is the same as $q^2=1$,
\begin{equation}\label{10.106}
\Upsilon=C\,\exp(-s),
\end{equation}
where $C$ is a constant. From this, one arrives at
\begin{equation}\label{10.107}
\mathcal{V}=\frac{\mathcal{A}}{r},
\end{equation}
where
$\mathcal{A}$ is another constant. Defining
\begin{align}\label{10.108}
a&:=-\frac{2\,\mathcal{A}}{J},\\
\label{10.109} b&:=1-\mathcal{E},
\end{align}
one arrives at
\begin{equation}\label{10.110}
\cosh^2 w\,\left[1+\left(\frac{\rd w}{\rd\alpha}\right)^2\right]=
(b-a\,\tanh w)^{-2},
\end{equation}
with
\begin{equation}\label{10.111}
w_0:=\tanh^{-1}\frac{a}{b},
\end{equation}
the differential equation is rewritten as
\begin{equation}\label{10.112}
\rd\alpha=\pm\frac{\sqrt{b^2-a^2}\,\cosh(w-w_0)\,\rd w}{\sqrt{1-(b^2-a^2)\,\cosh^2(w-w_0)}},
\end{equation}
the solution to which is
\begin{equation}\label{10.113}
\cos(\alpha-\alpha_0)=\sqrt{\frac{b^2-a^2}{1+a^2-b^2}}\,\sinh(w-w_0).
\end{equation}
Note that the case $a>b$ can also be handled with the same relation, using
\begin{equation}\label{10.114}
w_0=-\ri\,\frac{\pi}{2}+w'_0.
\end{equation}
Equation (\ref{10.113}) shows that $w$ is a function of $\alpha$
with the period $2\,\pi$, that is
\begin{equation}\label{10.115}
\mathcal{T}=2\,\pi.
\end{equation}
So for $c=0$, the orbits are closed.
\section{The additional symmetry}
In the case of commutative space, the fact that the bounded orbits of
the Kepler problem are closed is related to the fact that there is
an additional constant of motion in the Kepler problem, the so called
Laplace-Runge-Lenz vector, which points to the (fixed) periapsis
of the orbit. Here too, such an additional constant of motion happens to exist.
Defining
\begin{align}\label{10.116}
\mathfrak{p}&:=\cos\frac{k}{2},\\
\label{10.117} \bm{\bm{\varpi}}&:=\frac{2\,\bm{k}}{k}\,\sin\frac{k}{2},\\
\label{10.118} \bm{\rho}&:=\bm{x}-\frac{\bm{J}}{2},
\end{align}
one arrives at
\begin{align}\label{10.119}
\mathfrak{p}\,\bm{J}&=\bm{\rho}\times\bm{\varpi},\\
\label{10.120} \bm{x}&=\bm{\rho}+\frac{\bm{J}}{2},\\
\label{10.121} r^2&=\bm{x}\cdot\bm{x},\nonumber\\
&=\bm{\rho}\cdot\bm{\rho}+\frac{\bm{J}\cdot\bm{J}}{4}.
\end{align}
Using (\ref{10.15}) to (\ref{10.20}), with \cite{kfs}
\begin{equation}\label{10.0122}
x_a{}^b=\frac{k}{2}\,\cot\frac{k}{2}\,\delta_a^b+\left(1-\frac{k}{2}\,\cot\frac{k}{2}\right)\,
\frac{k_a\,k^b}{k^2}+\frac{1}{2}\,\varepsilon_a{}^{b\,c}\,k_c,
\end{equation}
one arrives at the following Poisson brackets.
\begin{align}\label{10.123}
\{x_a,\mathfrak{p}\}&=-\frac{\varpi_a}{4}.\\
\label{10.124}
\{x_a,\varpi_b\}&=\mathfrak{p}\,\delta_{a\,b}+\frac{1}{2}\,\varepsilon^c{}_{a\,b}\,\varpi_c.
\end{align}
One has
\begin{align}\label{10.125}
K&=\frac{4\,(1-\mathfrak{p})}{m},\\
\label{10.126} \bm{\varpi}\cdot\bm{\varpi}&=2\,m\,K\,\left(1-\frac{m\,K}{8}\right),\\
\label{10.127} H&=K+V,
\end{align}
where $V$ is a function of $r$. It is then seen that
\begin{align}\label{10.128}
\{\bm{\rho},H\}&=\frac{\bm{\varpi}}{m},\\
\label{10.129} r\,\{r,H\}&=\frac{\bm{\rho}\cdot\bm{\varpi}}{m},\\
\label{10.130} \{V,H\}&=\frac{1}{r}\,\frac{\rd V}{\rd r}\,\frac{\bm{\rho}\cdot\bm{\varpi}}{m},\\
\label{10.131} \{\bm{\rho}\,V,H\}&=\frac{1}{r}\,\frac{\rd V}{\rd r}\,
\frac{(\bm{\rho}\cdot\bm{\varpi})\,\bm{\rho}}{m}+V\,\frac{\bm{\varpi}}{m}.
\end{align}
Also,
\begin{align}\label{10.132}
\{\bm{\varpi},H\}&=-\frac{1}{r}\,\frac{\rd V}{\rd r}\,
\left(\mathfrak{p}\,\bm{\rho}-\frac{\bm{J}\times\bm{\varpi}}{4}\,\right),\\
\label{10.133} \{\bm{\varpi}\times\bm{J},H\}&=-\frac{1}{r}\,\frac{\rd V}{\rd r}\,
[(\bm{\rho}\cdot\bm{\varpi})\,\bm{\rho}-r^2\,\bm{\varpi}].
\end{align}
So, defining
\begin{equation}\label{10.134}
\bm{M}:=\frac{\bm{\varpi}\times\bm{J}}{m}+V\,\bm{\rho},
\end{equation}
one arrives at
\begin{equation}\label{10.135}
\{\bm{M},H\}=\left(r\,\frac{\rd V}{\rd r}+V\right)\,\frac{\bm{\varpi}}{m},
\end{equation}
which means that for the Kepler problem $\bm{M}$ is constant:
\begin{equation}\label{10.136}
\{\bm{M},H\}=0,\qquad V\propto r^{-1}.
\end{equation}
$\bm{M}$ is the analog of the Laplace-Runge-Lenz for the noncommutative case.
It can be seen that it fulfils algebraic relations similar to that of
the commutative case, namely
\begin{align}\label{10.137}
\bm{M}\cdot\bm{J}&=0,\\
\label{10.138} \bm{M}\cdot\bm{M}&=A^2+\left(\frac{2\,H}{m}
-\frac{H^2}{4}\right)\,(\bm{J}\cdot\bm{J}),
\end{align}
where
\begin{equation}\label{10.139}
V(r)=-\frac{A}{r}.
\end{equation}
Finally, the Poisson brackets between the constants
$\bm{J}$ and $\bm{M}$ can be calculated to arrive at
\begin{align}\label{10.140}
\{J_a,J_b\}&=\varepsilon^c{}_{a\,b}\,J_c,\\
\label{10.141} \{J_a,M_b\}&=\varepsilon^c{}_{a\,b}\,M_c,\\
\label{10.142} \{M_a,J_b\}&=\varepsilon^c{}_{a\,b}\,\left(-\frac{2\,H}{m}+\frac{H^2}{4}\right)J_c,
\end{align}
showing that the algebra of the constants of motion is the same as that of
the commutative case, with the simple replacement of
\begin{equation}\label{10.143}
H\to H-\frac{m\,H^2}{8}.
\end{equation}
\section{Concluding remarks}
The classical mechanics of central force problems in a noncommutative
three-dimensional space was investigated, where the Poisson structure of
the space coordinates is that of the SU(2) Lie algebra.
In particular, the aim was to find all of the potential energies
for which all of the bounded orbits are closed. A differential equation
was obtained for potential energies for which all nearly circular orbits
are closed. It was shown that among the solutions to that equation,
it is only the potential energy proportional to the inverse of the distance from
the origin (the potential energy corresponding to the Kepler problem)
for which all of the bounded orbits (not necessarily nearly circular)
are closed. As in the commutative case, this property of the Kepler
problem is related to an additional symmetry (the dynamic symmetry
leading to the conservation of the Laplace-Runge-Lenz vector), an analog
conserved quantity was searched in the noncommutative case here.
It was shown that there exists in fact a conserved vector (in addition
to the angular momentum). It was shown that it satisfies algebraic 
relations involving the angular momentum and the Hamiltonian, so 
that just similar to the commutative case it contains only 
one additional independent conserved quantity. The algebraic structure 
of the conserved quantities was also worked out and it was shown that 
the resulting algebra is similar to that of the commutative-space analog.
\\[\baselineskip]
\textbf{Acknowledgement}:  This work is
supported by the Research Council of the Alzahra University.

\end{document}